# Mobility Support for Cellular Connected Unmanned Aerial Vehicles: Performance and Analysis

Sebastian Euler, Helka-Liina Maattanen, Xingqin Lin, Zhenhua Zou, Mattias Bergström, and Jonas Sedin

Ericsson

Contact: sebastian.euler@ericsson.com

*Abstract*— Beyond visual line-of-sight connectivity is key for use cases of unmanned aerial vehicles (UAVs) such as package delivery, infrastructure inspection, and rescue missions. Cellular networks stand ready to support flying UAVs by providing wide-area, quality, and secure connectivity for UAV operations. Ensuring reliable connections in the presence of UAV movements is important for safety control and operations of UAVs. With increasing height above the ground, the radio environment changes. Using terrestrial cellular networks to provide connectivity to the UAVs moving in the sky may face new challenges. In this article, we share some of our findings in mobility support for cellular connected UAVs. We first identify how the radio environment changes with altitude and analyze the corresponding implications on mobility performance. We then present evaluation results to shed light on the mobility performance of cellular connected UAVs. We also discuss potential enhancements for improving mobility performance in the sky.

## I. INTRODUCTION

The operation of unmanned aerial vehicles (UAVs, a.k.a. drones) needs wireless connectivity for communications between UAVs and ground control systems, between UAVs themselves, and between UAVs and air traffic management systems [1]. The term UAV encompasses a wide range of aerial vehicles with vastly different sizes, weights, speeds, and flying altitudes. In this paper, we focus on low altitude small UAVs as described in Part 107 of the Federal Aviation Administration (FAA) regulations [2], with a weight below 55 pounds, a maximum speed of 100 miles per hour, and a maximum flying altitude of 400 feet above ground level (AGL) or within 400 feet of a structure if higher than 400 feet AGL.

Beyond visual line-of-sight (LOS) connectivity is key for use cases of low altitude small UAVs such as package delivery, infrastructure inspection, and rescue missions [3]. Cellular networks stand ready to support flying UAVs by providing wide-area, quality, and secure connectivity for UAV operations [4],[5]. The reliable support of mobile connections is one of the distinguishing features of cellular networks and why mobile operators can command higher cellular subscription fees than other forms of telephony and data access [6],[7]. Ensuring a reliable connection to UAV user equipment (UE) in the presence of UAV movement is important for safety control and operations of the UAV.

With increasing height above the ground, the radio environment changes [8]. Using existing mobile networks to provide connectivity to the UAVs moving in the sky may face new challenges. **Figure 1** gives an illustration of a cellular connected UAV UE. The UAV is flying well above the base station (BS) antenna height. Due to LOS propagation conditions, the UAV UE may suffer from higher downlink interference from the neighbor cells while the uplink signal from the UAV UE may increase interference in the neighbor cells compared to terrestrial UEs. Further, the BS antenna is usually tilted downwards by a few degrees, and the radiation pattern has one large main lobe covering the cell area. This optimizes coverage on the ground but leaves UAV UEs flying in the sky to be served by the antenna sidelobes. As a result, the strongest signal may come from a faraway BS that may be chosen by the UAV UE as its serving BS. In the example shown in **Figure 1**, the UAV UE moving horizontally from left to right might receive the strongest signal from the BSs in the order 1, 3, 2, and 1. The coverage areas of the sidelobes may be small and the signals at the edges of the antenna side lobes may drop sharply when the UAV is moving, due to deep antenna nulls.

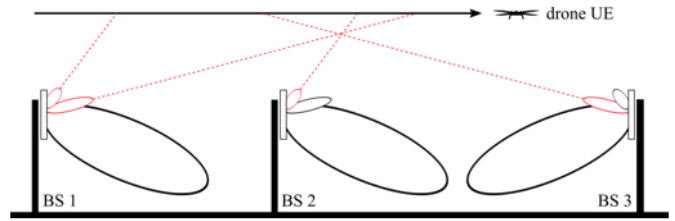

Figure 1: An illustration of a cellular connected UAV UE

The third-generation partnership project (3GPP), which is a global collaboration between groups of telecommunication associations for developing and maintaining system specifications for mobile technologies, dedicated a significant effort during its Release 15 to study the potential of terrestrial cellular networks, particularly Long-Term Evolution (LTE), for connecting UAVs [9]. This study was completed in December 2017 and the outcomes are documented in the 3GPP technical report TR 36.777 [10]. Among others, a key objective during the study was to investigate the mobility performance of UAV UEs, including to identify if robustness in handover (HO) signaling can be achieved and if enhancements in terms of cell selection and HO efficiency are needed. As shown by simulation and field trial results submitted to 3GPP, in some scenarios the mobility performance of UAV UEs is worse compared to terrestrial UEs [10]. Based on this study, 3GPP concluded that LTE networks can serve UAVs, but there may be challenges including mobility. With the completion of the study item, 3GPP started a follow-up work item [11] to advance LTE technologies including mobility enhancements to provide more efficient and reliable cellular connectivity to aerial vehicles.

Mobility management for airborne UAVs is an interesting field that deserves further investigation. While the changes in the radio environment with height are well known and generally taken into account in current studies [12],[13],[14], the implications for moving UEs and for the established mobility procedures are not yet well understood. In this article, we share some of our findings concerning mobility support for cellular connected UAVs. We present evaluation results to shed light on the mobility performance of LTE connected UAVs and discuss potential enhancements for improving mobility performance in the sky in existing networks that have been optimized for terrestrial coverage.

## II. PERFORMANCE METRICS AND MODELING METHODOLOGY

Important metrics for the evaluation of mobility performance are the numbers of successful and failed HOs, the number of radio link failures (RLF), and the rate of ping-pong HOs. HO and RLF are defined procedures in the 3GPP LTE radio resource control (RRC) specification [15]. HO failure and ping-pong HO are additional mobility performance metrics, which were defined for performance evaluation purposes during an earlier 3GPP study on mobility enhancements for heterogeneous networks and captured in the 3GPP technical report TR 36.839 [16]. In this paper, we adopt these metrics and the modeling methodology for the evaluation and analysis of LTE connected UAV UEs. The two key aspects in the modeling are the radio link monitoring (RLM) process and the HO process, which are illustrated in **Figure 2**.

### A. Radio Link Monitoring Process

A RLF occurs when the UE cannot establish or maintain a stable connection to the serving cell. According to [15], the UE declares RLF upon indication from the radio link control (RLC) layer that the maximum number of retransmissions has been reached, or upon expiry of the timer T310 that is started when physical layer problems are detected, or upon indication from medium access control layer on random access problems.

For mobility evaluations, RLF triggered upon the expiry of T310 is considered. In the RLM process, the UE periodically computes a channel quality indicator (CQI) by evaluating the signal quality of a reference signal in the physical downlink control channel (PDCCH). If the CQI drops lower than a threshold Qout, it is considered "out-of-sync." Higher layers count subsequent out-of-sync indications. If a maximum number of consecutive out-of-sync indications (denoted by N310) is reached, the UE starts timer T310, whose expiry would trigger RLF. While the timer T310 is running, the UE periodically evaluates the signal quality. If it recovers, the UE stops the timer, does not declare RLF and maintains the RRC connection. If the quality of PDCCH does not improve while the timer is running, the UE declares RLF upon the expiry of T310. If RLF is declared, the UE either tries to re-establish the RRC connection or goes back to idle mode and starts the cell search procedure in order to establish a new RRC connection.

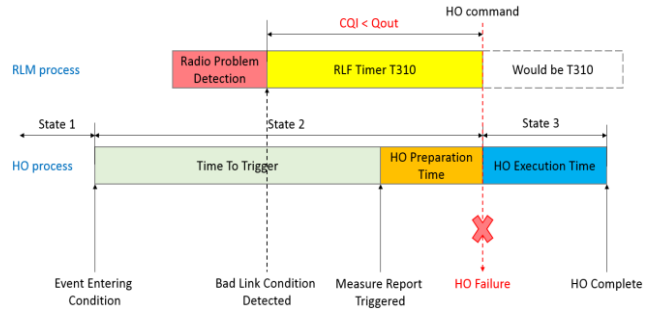

**Figure 2: Illustration of RLM and HO processes [16]**

### B. Handover Process

HO failure (HOF) is defined in the context of the HO process, which is initiated by measurements of the reference signals fulfilling certain conditions, so-called events, as illustrated in **Figure 2**. According to [16], the HO process can be divided into the following three states.

- **State 1:** Before the event (e.g. event A3 indicating that a neighbor becomes offset better than the serving cell) entering condition is satisfied, i.e., before the actual HO procedure has started.
- **State 2:** After the event entering condition is satisfied, but before the HO command is received by the UE, i.e., while the UE is waiting for and expecting the HO command.
- **State 3:** After the HO command is received by the UE, but before the HO complete is sent by the UE, i.e., during the HO execution.

In this process, a HOF is counted if any of the following occurs:

- If RLF occurs in State 2, i.e., if the quality of the serving cell drops too fast before the HO to the target cell can be executed.
- If the timer T310 (at whose expiration the UE would declare RLF) is running when the HO command would be received, i.e., if the UE cannot receive the HO command from the source cell due to the poor link quality.
- If a PDCCH failure (using the same criterion as for starting the timer T310) occurs in State 3, i.e., if the target cell signal quality turns out to be too low to establish a connection.

A ping-pong HO (PP) is defined as a HO that is followed by another HO back to the original cell, occurring within some time window t. Throughout this study, we have used t = 1s in accordance with the definition used in [16].

## III. EVALUATION ASSUMPTIONS

We use a full system-level simulator, capable of simulating a large number of UEs moving through a cellular network with a regular hexagonal cell layout. The evaluation assumptions follow the 3GPP study item on enhanced LTE support for aerial vehicles [10]. For ease of reference, we summarize in **Table 1**

Table 1: Evaluation Assumptions

| Parameter | Value | |
|---|---|---|
| | UMa | RMa |
| Cell layout | Hexagonal grid, 19 sites, 3 sectors per site | |
| Inter-site distance (ISD) | 500 m | 1732 m |
| BS antenna height | 25 m | 35 m |
| Carrier frequency | 2 GHz | 700 MHz |
| BS antenna pattern | As specified in TR 36.873 [17], with (M, N, P) = (8, 1, 2) where M denotes the number of rows in the array, N denotes the number of columns in the array, P denotes polarization | |
| BS antenna downtilt angle | 10 degrees | 6 degrees |
| UE density | 15 UAV UEs per cell, no terrestrial UEs | |
| UE height | {0, 50, 100, 300} m | |
| UE speed | {3, 30, 60, 160} km/h | |
| Traffic model | {Full buffer, FTP} in downlink | |
| Event A3 offset | 2 dB | |
| Time-to-trigger (TTT) | 160 ms | |
| T310 | 1 s | |

the key evaluation assumptions pertinent to mobility performance studied in this paper.

We study both urban-macro (UMa) and rural-macro (RMa) scenarios. In each simulation, all UEs are placed at the same altitude and have the same speed, but random starting points and directions in the x-y plane. Speed and height stay constant during the simulation. Four different heights (ground level, 50 m, 100 m, 300 m) and four different UE speeds (3 km/h, 30 km/h, 60 km/h, 160 km/h) are simulated. The simulations are repeated with two different traffic models: a full-buffer model, where the network is used to full capacity (i.e. 100% resource utilization), and a file transfer protocol (FTP) traffic model with the parameters (FTP object size and reading time) chosen such that an intermediate level of resource utilization is realized.

In the simulations, all failure types (HOF and RLF) are logged separately for each state described in Section II. This provides rich information on the states in which the failures occur as well as the causes. Note that in the results presented below, HOFs and RLFs are counted in a mutually exclusive way: if a RLF occurs in State 2 of the HO process, it is counted as a HOF only and not logged again as part of the final RLF statistics. For the HO rates presented, they include both the successful and failed HOs.

## IV. BASELINE MOBILITY RESULTS

### A. UMa Scenario with Full-buffer Traffic

In this section, we study the UMa scenario with full-buffer traffic, which is chosen as a baseline. It should be noted that this scenario represents sort of a worst case because of its high UE density per unit area (compared to the RMa scenario, which features the same number of UEs, but within significantly larger cells) and high network load (100% resource utilization), and the resulting high levels of interference.

**Figure 3**(a) and **Figure 3**(b) show the HO rate and RLF rate for the four simulated UE heights and four speed settings. In line with expectations, the HO rate is increasing with speed, as a faster UE passes through more cells than a slower UE during the same time window. An interesting observation is that HO rate and RLF rate are negatively correlated: With increasing height, the UEs perform fewer HOs, while the RLF rate increases strongly. The key takeaway is that UAV UEs often go into RLF instead of initiating the HO process. When the serving cell quality decreases rapidly, the out-of-sync indications due to low PDCCH quality start the timer T310, which expires before a measurement event A3 is triggered, meaning that no candidate cell is above the threshold.

The above phenomenon can be illustrated by **Figure 3**(c), which shows an example of simulated Reference Signal Received Power (RSRP) and Signal-to-Interference-plus-Noise Ratio (SINR) traces for a UAV UE moving for 10 s at a height of 300 m and with a speed of 30 km/h. Each colored line in the RSRP subplot corresponds to the RSRP measurements of one cell. The vertical dark green dashed line at the beginning of the simulation marks cell selection of the cell with the corresponding color. After 3 s, the serving cell RSRP begins to drop. After 5 s, the RSRPs from some neighboring cells become stronger than that of the serving cell. After 6 s, the RSRPs from all the neighboring cells become stronger than that of the serving cell. However, the RSRPs of the neighbor cells are all at about the same level and stay relatively low. None of them is at least 3 dB better than the serving cell, which is the A3 threshold in this particular simulation and thus the condition to trigger a measurement report. After 7 s, the UE declares RLF (marked by the vertical red dashed line) due to poor serving cell SINR, without having even sent a measurement report, which would have been a prerequisite for initiating a HO.

An even more extreme example mobility trace is shown in **Figure 3**(d). Here the UE starts in the light green cell, before being successfully handed over to the dark green cell after about 3.5 s. At about 5 s, however, the UE moves through a null between two sidelobes of the same BS antenna. The RSRP of the serving cell drops sharply, about 10 dB within only one second (corresponding to a distance of about 8 m at a speed of 30 km/h). At the bottom of the dip the UE inevitably declares RLF.

These examples illustrate the two main challenges for providing mobility support for UAV UEs using existing terrestrial cellular networks: the stability of the signal strength of the serving cell, and the interference fluctuation. Sudden drops in signal strength due to the UE moving through antenna nulls between sidelobes might lead to frequent RLFs, because the default HO procedure may simply be too slow to be successfully executed. We can see further that the gaps between the serving cell RSRP and the neighbor cell RSRPs are small.

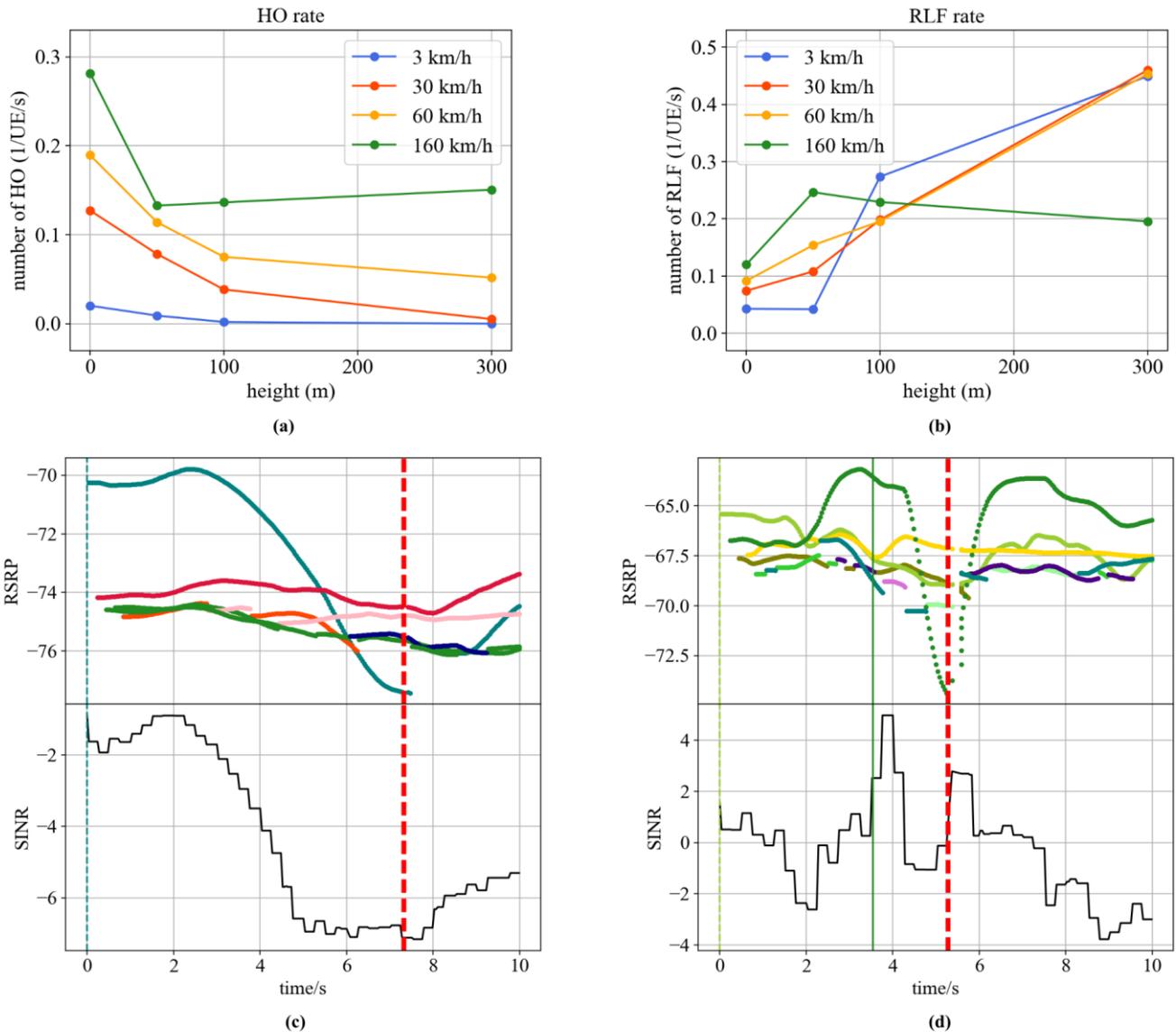

Figure 3: UMa scenario with full-buffer traffic: (a) HO rate, (b) RLF rate, (c) and (d) two example mobility traces

The strong interference from neighbor cells makes the serving cell SINR stay relatively low throughout.

*B. RMa Scenario with FTP Traffic*

In this section, we study the RMa scenario with an FTP traffic model, which is more realistic than full-buffer traffic. Here the simulated cell area is larger by a factor of ~10 (ISD of 1732 m vs. 500 m for the UMa case), which translates to a correspondingly lower UE density per unit area. The FTP parameters are chosen to result in an intermediate network load level. Because the parameters are kept constant for all speed and height settings, load levels may however change between simulation cases, since the needed radio resources vary with speed and height. As an example, consider the resource utilization for a UE speed of 30 km/h at the 4 simulated heights. On the ground, the chosen settings realize a resource utilization level below 10%. At 50 m and 300 m, the lower spectral efficiencies lead to a higher resource utilization of around 30%.

The resulting HO and RLF rates are shown in **Figure 4**(a) and **Figure 4**(b). Again, there is a negative correlation between HO rate and RLF rate. In contrast to the UMa scenario with full-buffer traffic (and with the notable exception of the 100 m height setting), we see quite good mobility performance. Here, the HO rates do not decrease with height, and the RLF rates are low, especially at 300 m height. This is due to lower UE density per unit area, lower network load, and larger cell size, resulting in much lower interference levels.

Another aspect of the mobility performance is the HOF ratio, i.e., the fraction of attempted HOs that fail, either because the quality of the serving cell drops too fast such that the measurement reports or the HO commands are lost during transmission, or because the quality of the target cell turns out to be not good enough such that the UE cannot establish a new connection (see Section II). The fraction of failed HOs for the RMa scenario with FTP traffic simulation is shown in

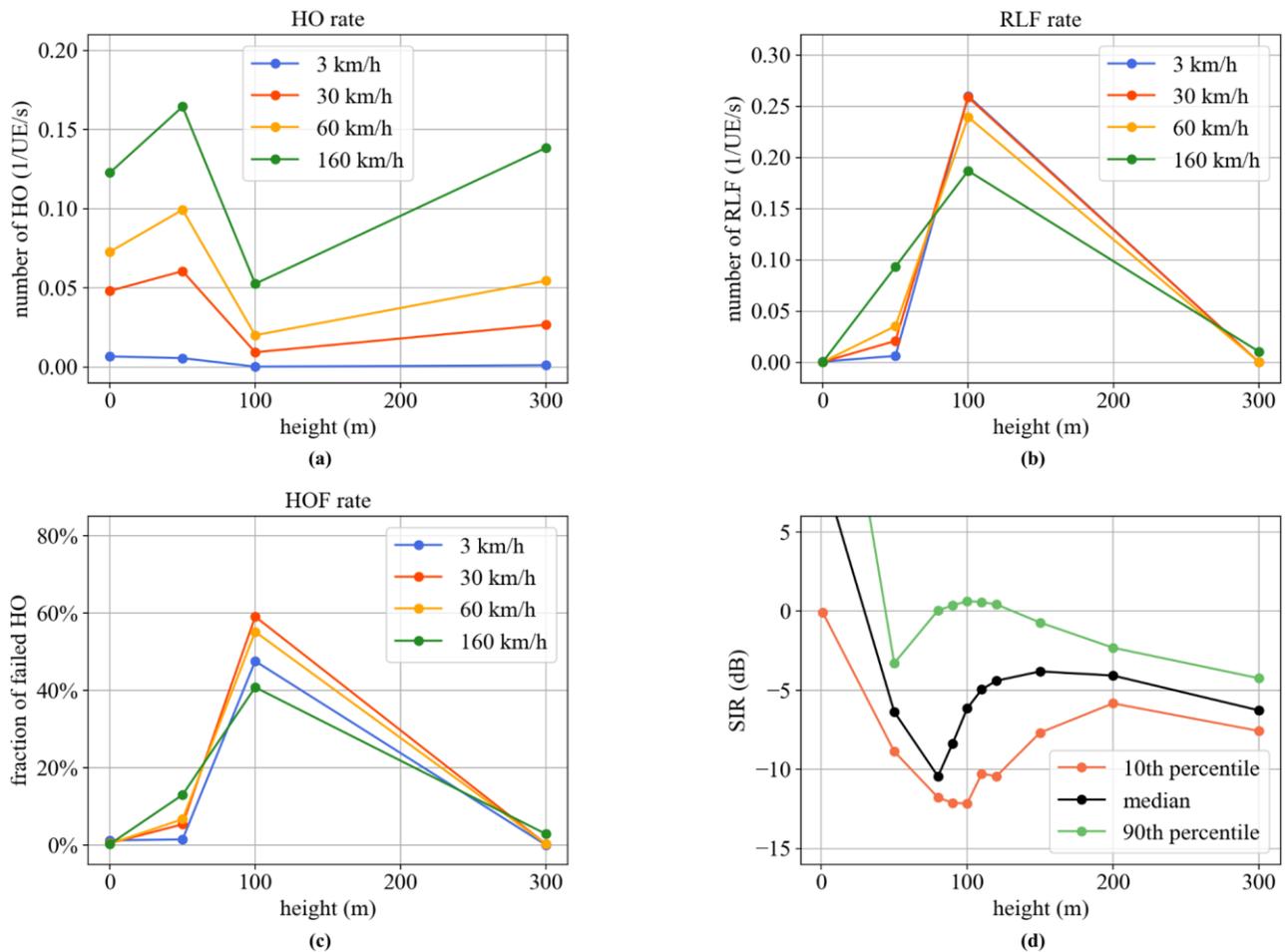

**Figure 4: RMa scenario, FTP traffic: (a) HO rate, (b) RLF rate, (c) HOF rate, (d) 10[th] percentile, median, and 90[th] percentile SIR**

**Figure 4**(c). Again, we observe good mobility performance with low HOF rates at all heights except at 100 m, where between 40% and 60% (depending on the UE speed) of all attempted HOs (which are already fewer than at other altitudes according to **Figure 4**(a)) fail.

The anomalous phenomenon at 100 m height, as indicated by the poor HO, RLF, and HOF performance (shown in **Figure 4**(a-c), respectively), is reflected by the somewhat surprising observation that at 100 m height we see an excessively high resource utilization level (above 80%). To understand the anomalous behavior at 100 m height, we examine the distributions of signal-to-interference ratio (SIR) values for the RMa scenario. **Figure 4**(d) shows the 10[th] percentile, median, and 90[th] percentile SIR as a function of height. It can be clearly seen that there exist areas with poor SIR around 100 m height with large variance. At larger heights, the variance becomes smaller, and the 10[th] percentile SIR is much higher than the counterpart at 100 m height.

It should be noted that the SIR distribution at a certain height depends on the network deployment and BS antenna patterns. The existence of large regions with very low SIR at a specific height might be a consequence of the regular hexagonal deployment with identical BS antenna patterns that are used in this simulation. Nonetheless, even in more irregular and thus more realistic deployments we expect the existence of regions in the sky with poor radio conditions, where UAV UEs might experience disproportionately high rates of RLF and HOF.

## V. MOBILITY RESULTS WITH SELECTED ENHANCEMENTS

The mobility results for at least the RMa scenario with FTP traffic (Section IV.B) are promising already with the legacy LTE HO mechanisms. Still, an important challenge – and a central point of the 3GPP work item [11] – was to introduce enhancements to the LTE mobility procedures with the potential to further improve the performance and to ensure reliable operations even in adverse conditions, such as in the UMa scenario studied in Section IV.A. One such possible enhancement is the coverage extension feature introduced for (but not exclusive to) LTE machine type communications (LTE-M) in Release 13, which allows UEs to operate in worse SINR conditions. By introducing repetition, the same error rate can be achieved at lower SINR. To investigate the potential of the LTE-M coverage extension feature, we evaluate the RMa scenario with FTP traffic at 50 m height again, but with the Qin/Qout thresholds lowered from the default values of -6 dB/-8 dB to -10 dB/-12 dB, respectively. This eliminates almost all RLF and HOF, at the cost of a slightly increased PP rate, as shown in **Figure 5**, as well as increased resource utilization.

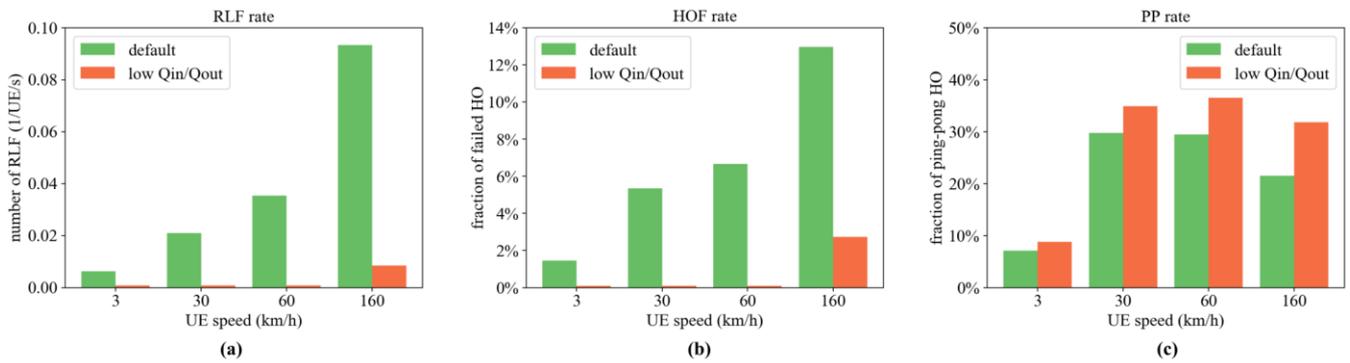

**Figure 5: Mobility performance at 50 m height with lower Qin/Qout: (a) RLF rate, (b) HOF rate, (c) PP rate**

## VI. CONCLUSIONS AND FUTURE WORK

The study in this paper suggests that the existing terrestrial LTE networks should be able to offer wide-area wireless connectivity with good mobility support to the initial deployment of a small number of UAVs. Higher UAV densities and/or more difficult radio environments might be challenging. Two main problems have been identified:

- When UAV UEs move through BS antenna sidelobe nulls, the default mobility procedures might be too slow to be successfully executed. UAV UEs might declare RLF before a HO to another cell can be completed.
- UAV UEs experience LOS propagation conditions to many neighbor cells, which results in comparably high interference levels. This makes it difficult to establish and maintain connection to the network, which might lead to increased RLF and HOF rates.

Potential solutions to both problems are being investigated. A possible solution to the latter might be the coverage extension feature introduced for LTE-M, which allows UEs to connect to the network under worse SINR conditions. In our simulations, with this enhancement, we were able to significantly reduce the rates of RLF and HOF, at the cost of a slightly increased PP rate and resource utilization. In addition, careful flight planning might be used to avoid regions of poor SINR in the sky. The positions of such regions could be determined by the network, for example by evaluating the measurement reports of UAV UEs. The network might then in turn inform UAV UEs about these regions, so that these UAVs could adapt their flight paths to avoid the coverage holes in the skies. A complementary approach is to reduce the amount of interference generated in the first place, e.g., by using directional antennas at the UE side.

To solve the first problem, one possible solution is to reduce the reaction time of the mobility procedures, e.g., by tuning the HO parameters. A more advanced solution is the introduction of a conditional HO procedure, where the network preemptively (without the UE having sent a measurement report) sends a HO command to the UE with an added condition. Upon fulfillment of this condition, the UE can immediately initiate the HO, without having to wait for the HO command.